# An Open-Source Framework for Measurement and Analysis of Nanoscale Ionic Transport

Yichao Wang[1], Munan Fang[1,2], Aziz Roshanbhai Lokhandwala[1,2], Siddhi Vinayak Pandey[1,2], Boya Radha [1,2]*

[1]Department of Physics and Astronomy, School of Natural Sciences, The University of Manchester, Manchester M13 9PL, United Kingdom

[2]National Graphene Institute & Photon Science Institute, The University of Manchester, Manchester M13 9PL, United Kingdom

*Correspondence to be addressed to: Boya Radha (email: radha.boya@manchester.ac.uk)

## Abstract

Nanofluidic systems exploit nanometre-scale confinement in channels and pores to regulate ionic transport, enabling functionalities such as osmotic energy harvesting and neuromorphic ionic memory. Studying such confined transport requires both precise electrical instrumentation and careful data analysis, yet, in practice, measurements are still often taken with vendor software, exported as files, and processed later in separate environments. In this work, we bring these steps together in a unified Python-based framework built around three interoperable graphical user interfaces (GUIs) for nanochannel, nanopore and memristor experiments. The framework is organised into two functional parts, measurement and analysis. On the measurement side, two GUIs drive Keithley Source Meters to run continuous voltage sweeps and user-defined memristive pulse sequences, while providing live plots, configuration management and controlled shutdown routines. On the analysis side, a dedicated nanochannel and nanopore GUI reads raw I-V datasets, applies unit-consistent processing, extracts conductance and ion mobility, evaluates selectivity and osmotic power, and is complemented by a web-based calculator that performs the same mobility analysis without a local Python installation. All three GUIs are implemented in Python/Tkinter with modular plotting and logging layers so that flexible control sequences and physics-based post-processing share a common data format, improving reproducibility, timing stability and day-to-day efficiency in nanofluidic and electronic device studies.

## 1. Introduction

Emerging from foundational microfluidic technologies[1,2], Ångström-scale and nanoscale fluidic conduits, including synthetic nanopores, slit-like nanochannels, and one-dimensional nanotubes, now serve as a unified platform for investigating transport phenomena under extreme confinement[3,4]. By mimicking the dimensional constraints of biological ion channels, these engineered systems replicate the exquisite ion selectivity essential for cellular signaling. As electrolyte solutions traverse these confined geometries, the interaction between ions and atomically smooth walls gives rise to unique transport phenomena, enabling diverse applications ranging from osmotic energy harvesting and precise molecular filtration to neuromorphic computing and ionic memory[5–7]. Representative device concepts for two such application classes, solid-state nanopores for DNA sequencing and nanochannel memristors for neuromorphic iontronics (Fig.1).

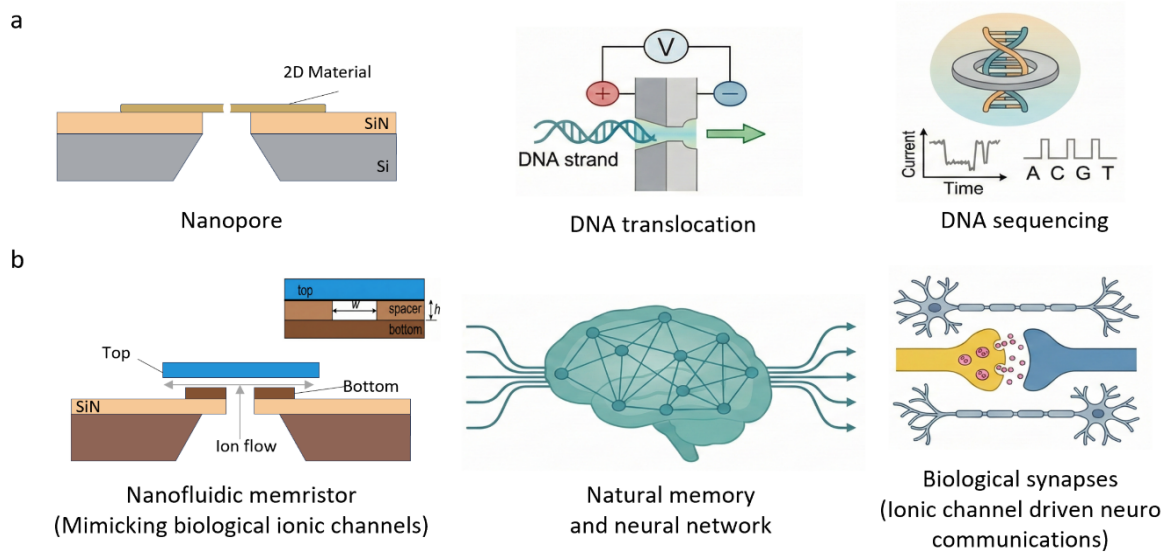

**Figure 1.** Representative nanoscale-fluidic device concepts linking functionality to application domains. a. Nanopores based on two-dimensional materials used for single-molecule DNA sequencing (device structure, nanopore capture, and current-trace-based base calling); b. Å-scale nanochannel memristors used for neuromorphic iontronics (fabricated slit channel, network-level abstraction, and biological synapse analogy).

Facilitated by rapid progress in both top-down nanofabrication and bottom-up self-assembly, it is now possible to create reproducible channels with atomic precision that closely mimic the dimensions of biological counterparts[6]. Key examples include van der Waals assembled two-dimensional slits, graphene and $MoS_2$ membranes, carbon nanotube conduits, and solid-state nanopores[8–10]. The combination of atomically thin walls and tunable surface chemistry in these systems enables the systematic exploration of size, geometry, and charge-dependent transport, leading to phenomena such as ionic rectification, nonlinear mobility, osmotic power generation, and beyond-steric ion selectivity[11,12]. Moreover, such Ångström-scale channels have revealed nonlinear transport behavior reminiscent of memristive dynamics, where ion accumulation and surface charge memory give rise to pinched I–V hysteresis[13].

Despite these advances in device fabrication, the complementary measurement and analysis workflows remain a bottleneck. On one hand, commercial instrument suites (e.g., BioLogic EC-Lab or Keithley KickStart) provide robust hardware control but operate as closed-source ecosystems designed for general electrochemistry rather than specialized nanofluidic physics[14]. On the other hand, community-developed analysis packages such as Transalyzer or OpenNanopore excel at offline event segmentation but typically lack direct hardware integration[15–17]. Even recent protocols demonstrating automated feedback control for nanopore fabrication[18], often rely on proprietary environments like LabVIEW, which limits their seamless integration with modern, open-source data science libraries.

Consequently, researchers rely on a disjointed pipeline: acquiring data via vendor software, manually exporting files, and performing post-processing in separate environments. For instance, characterizing transport in Å-scale channels typically requires switching between acquisition scripts and disconnected analysis tools such as Origin or MATLAB[3]. This fragmentation becomes particularly problematic in two scenarios: (i) when characterizing steady-state ionic transport, where continuous I–V sweeps under well-defined geometries and salinity gradients must be combined with device dimensions to extract membrane potentials, conductance, ion mobility, and osmotic power density from the same dataset; and (ii) when investigating memristive neuromorphic dynamics, which necessitates correlating programmable, sub-second voltage pulse sequences (e.g., spike-timing-dependent plasticity (STDP) protocols) with dynamic conductance modulation, state retention, and loop-area based power or power–frequency profiles derived from hysteretic I–V curves. Parallel to the evolution of genomics, where integrated computational pipelines became essential to manage increasing data complexity[19], the field of nanofluidics now necessitates a similar convergence of real-time instrument control and physics-based analytical modeling to ensure scalability and reproducibility.

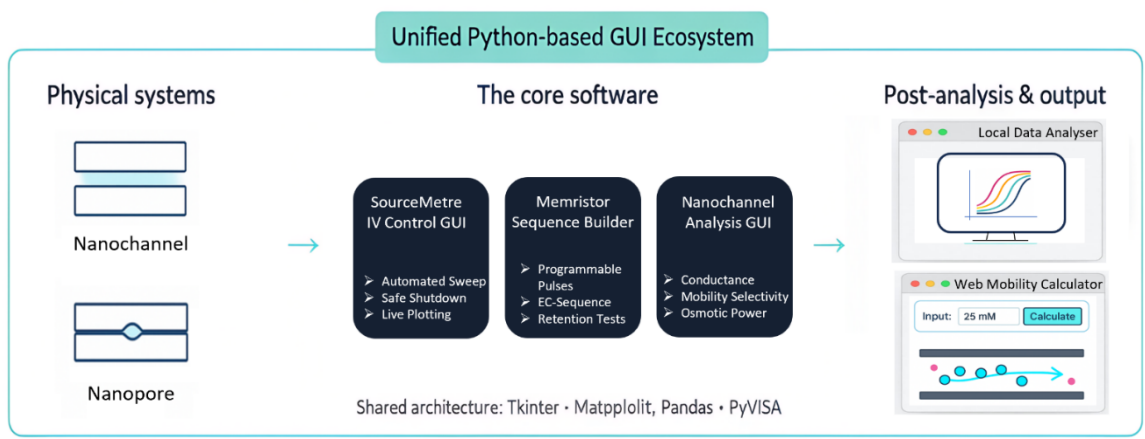

**Figure 2.** Graphical overview of the proposed GUI ecosystem for nanoscale transport experiments.

To bridge this gap, we developed a unified, Python-based graphical user interface (GUI) ecosystem that seamlessly integrates instrument control with domain-aware analysis. The framework provides an end-to-end environment spanning three interoperable modules: (i) a Measurement Control module that automates continuous and drift–diffusion I–V sweeps under user-specified geometries and concentration gradients, capturing the steady-state characteristics required for conductance, membrane-potential, mobility, and osmotic power-density extraction; (ii) a Sequence Builder tailored for memristive experiments, enabling the precise generation of complex synaptic learning protocols (e.g., paired-pulse, STDP) and logging the resulting hysteretic I–V loops so that dynamic conductance changes, state retention, and loop-area based power or power–frequency behavior can be quantified in a single workflow; and (iii) a Data Analysis module that directly imports experimental datasets from both acquisition paths to provide unit-consistent plotting and to compute ion mobilities, osmotic power, and memristive switching parameters (e.g., power versus conductance or frequency) using built-in physical models. Complementing these desktop applications, we also deploy a web-based version of the mobility calculator, offering an installation-free solution for rapid extraction of ion mobility and related transport parameters without the need for a local Python environment. By consolidating these functions into a single open-source framework, the system eliminates manual data transfer steps, improves timing precision, and standardizes parameter extraction for reproducible nanoscale transport research. A graphical overview of the overall workflow is provided in Fig. 2. The remainder of this paper is organized as follows: Section II reviews related work, Section III details the system and GUI architectures, Section IV presents experimental methods and validation, and Section V concludes the paper.

## 2. Related Works

### 2.1. General-purpose Measurement Frameworks

Experimental automation depends on software to control multiple instruments and to collect and organize the data. Open-source ecosystems such as QCoDeS[20] and PyMeasure[21] have become standard in nanoelectronics for defining complex, n-dimensional parameter sweeps. However, these frameworks operate primarily as "host-side orchestration engines" rather than real-time controllers. As noted by Riesebos et al., QCoDeS is not designed to control instruments in real time during execution[22], and Versluis et al. highlight that achieving low-latency feedback loops requires tighter hardware integration than general-purpose Python stacks typically offer[23]. In practice, Python interpreter latency and host-side polling further limit sub-second decision-making loops, which are essential for protecting sensitive nanofluidic devices during abrupt current excursions.

Commercial solutions such as Keithley KickStart offer user-friendly interfaces for I–V characterization, data logging, and multi-instrument management[14]. Yet these suites often act as closed-source "black boxes", limiting users to fixed workflows and licensing models that hamper integration into customized research pipelines. Because the internal logic is not user-extensible, it is not possible to embed custom physics models such as drift–diffusion, mobility extraction, or osmotic-power calculations directly into the acquisition loop[24,25].

Similarly, the LabVIEW ecosystem serves as a dominant platform for instrument control but faces challenges in modern software engineering. While LabVIEW provides extensive instrument driver support through the IVI standard[26,27], standard host-side Virtual Instruments (VIs) are non-deterministic, necessitating Real-Time (RT) or FPGA modules for precise timing[28]. Furthermore, surveys and industrial reviews highlight inherent difficulties in version control, cross-platform portability, and dynamic user interface scaling within legacy LabVIEW environments[29,30]. Because LabVIEW VIs are stored as binary files, they do not interact well with modern version control; changes to block diagrams cannot be inspected or merged in a readable way. In contrast, our framework is implemented in plain Python, which integrates cleanly with Git workflows and supports cross-platform use, structured data handling, and responsive near real-time control.

### 2.2. Nanopore and Nanochannel Analysis Tool

In the domain of nanopore sensing, analysis packages such as OpenNanopore (Python) and Transalyzer (MATLAB) are now well established and widely used for current–time trace analysis[15–17]. They are particularly effective for offline detection, segmentation, and statistical treatment of transient blockade events. However, these tools are typically run separately from the acquisition software, so data must first be recorded with vendor programs and then exported manually for analysis. This acquire–export–analyze workflow makes it difficult to apply a single, consistent pipeline when comparing rectification ratios, ionic mobility, or conductance–concentration curves across multiple devices and experiments. More recent GUI-based platforms such as Nanolyzer improve interactivity for event-driven analysis and offer a more streamlined user experience[31], but they remain focused on sequencing-style applications rather than the continuous, steady-state I–V measurements needed to extract ionic conductance and mobility in nanochannels[5,6].

While advanced setups utilizing FPGA–LabVIEW "ping-pong" architectures have achieved sub-microsecond feedback for molecular manipulation[32,33], these represent bespoke, highly specialized engineering solutions. Therefore, we developed a GUI that unifies continuous I–V acquisition with physics-based transport analysis, including drift–diffusion modeling and osmotic power estimation, enabling an end-to-end, interactive loop without switching between separate tools.

### 2.3. Sequence Controller for Memristor

The intersection of nanofluidics and neuromorphic computing has spurred interest in memristive ion transport. Recent studies have demonstrated that Ångström-scale channels exhibit volatile and non-volatile memory states[13], as well as synaptic plasticity rules such as spike-timing-dependent plasticity (STDP)[34,35]. Investigating these phenomena requires precise, programmable voltage pulse trains.

However, instrumentation for such experiments remains fragmented. Most memristor studies rely on bespoke scripts to execute paired-pulse facilitation (PPF) or STDP protocols, which complicates reproducibility and sharing of pulse recipes. General automation frameworks lack specialized sequencers for defining synaptic timing rules, while standard electrochemistry software typically supports only rigid waveforms such as cyclic voltammetry rather than the asymmetric, multi-stage pulse trains required for neuromorphic characterization. The EC-Sequence module presented in this work fills this gap by enabling researchers to design, visualize, and stream synchronized pulse protocols directly to SourceMeters, thereby unifying ionic transport and memristive dynamics studies within a single platform.

## 3. Methodology

### 3.1. System Architecture and Overview

This work presents an interoperable toolchain of three Python GUIs for nanochannel experiments, built with Tkinter, Pandas, and Matplotlib, with mplcursors for interactive visualisation, and PyVISA for instrument communication in the measurement modules. The system separates measurement control (Nanochannel Measurement Control GUI and EC-Sequence Control GUI) from post-run analysis (Nanochannel Data Analysis GUI) while using shared data formats and plotting standards to enable direct transfer of datasets and parameters between modules, as summarised in Fig. 3.

Each GUI operates independently yet follows a common architectural pattern comprising four principal modules, with functional emphasis varying across applications: (i) User Interface Layer, which handles parameter inputs, button callbacks, and interactive displays (implemented in all GUIs); (ii) Acquisition and Control Layer, which communicates with measurement instruments such as the Keithley 2600-series SourceMeter via VISA/SCPI commands (implemented in the Measurement GUI and EC-Sequence GUI, not required in the Analysis GUI); (iii) Computation and Analysis Layer, which implements numerical fitting, conductance extraction, drift–diffusion and mobility computation, and region-of-interest (ROI) processing (core of the Analysis GUI, secondary in the other two); and (iv) Visualization and Export Layer, which generates real-time plots, interactive cross-point detection, and exports data and figures in CSV and PNG formats (common to all GUIs).

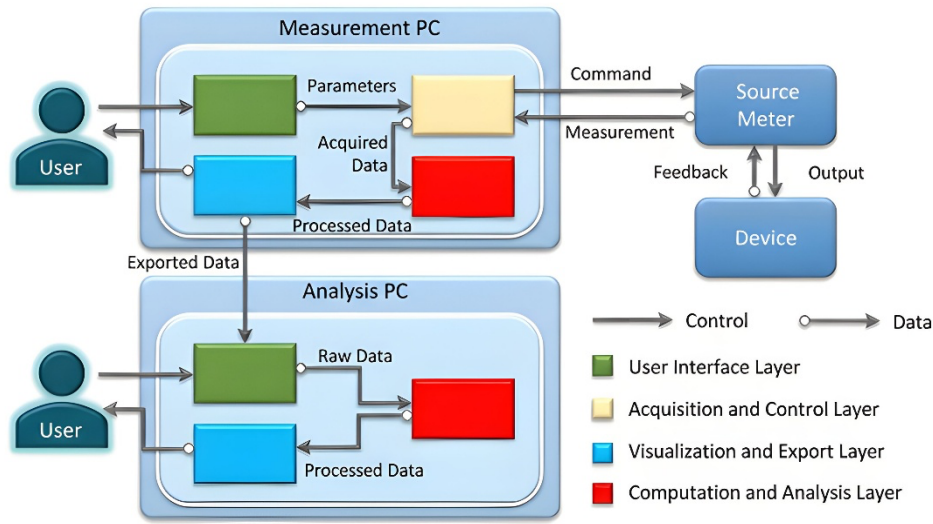

**Figure 3**. Overall architecture of the unified acquisition--analysis system.

This layered structure ensures consistency in unit handling, real-time responsiveness during measurement, and reproducibility across experiments. The following subsections describe the three primary GUIs that together constitute the integrated experimental platform.

### 3.2. Nanochannel Measurement Control GUI

The nanochannel measurement control GUI directly interfaces with a Keithley 2600-series SourceMeter via the PyVISA layer using standard SCPI/TSP command sets. All voltage sweep parameters, including start and stop values, step size, dwell time per step, compliance current, and number of sweep cycles, are user-defined through dedicated Tkinter entry fields. Once a run is initiated, the controller constructs a sweep vector in Python and transmits each voltage setpoint sequentially to the SourceMeter, enforcing the configured dwell interval through synchronized delays.

To ensure non-blocking operation, the system adopts a producer–consumer threading model (Fig. 4). The acquisition thread continuously queries the device buffer for real-time voltage and current readings, then places (V, I) tuples into a bounded queue. The main Tk thread runs a lightweight "LivePlotService" consumer that drains this queue at up to 20 frames s$^{-1}$, updating a Matplotlib canvas embedded in the GUI. This architecture maintains UI responsiveness while preserving sub-25 ms timing fidelity between consecutive samples.

All raw measurements, including timestamps, channel identifiers, and configuration metadata, are logged in parallel by a dedicated writer process to standardized CSV files for post-run analysis. Automatic routines monitor compliance events in real time, detect overcurrent or overvoltage conditions, and either adjust the measurement range or terminate the sweep if thresholds are exceeded. The controller continuously verifies the output state and instrument health through periodic status queries, ensuring recovery from communication or buffer faults.

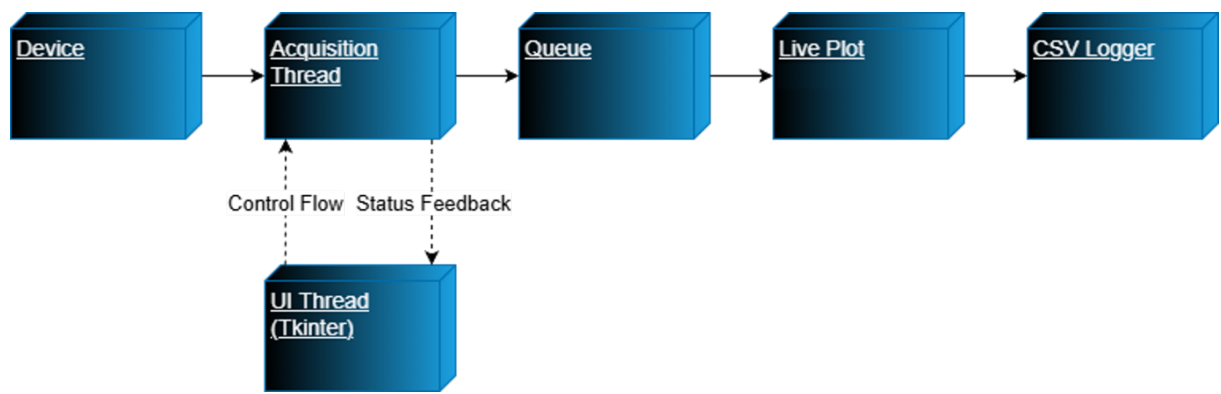

Figure 4. Threading and data-flow structure of the Keithley control module, showing asynchronous data acquisition, queue-based communication, and Tkinter-controlled visualization.

Built-in safety routines enforce compliance limits, manage range transitions, and coordinate multilevel shutdown sequences (Fig. 5). The output channel is first ramped down to zero, then explicitly disabled through the command smua.source.output = smua.output_off, followed by a VISA resource reset. These layered protections prevent damage to both the device under test and the SourceMeter hardware, even under abnormal load or user-interrupted conditions. The resulting datasets can be directly imported into the Analysis GUI for conductance fitting, G--V conversion, and drift--diffusion calculations.

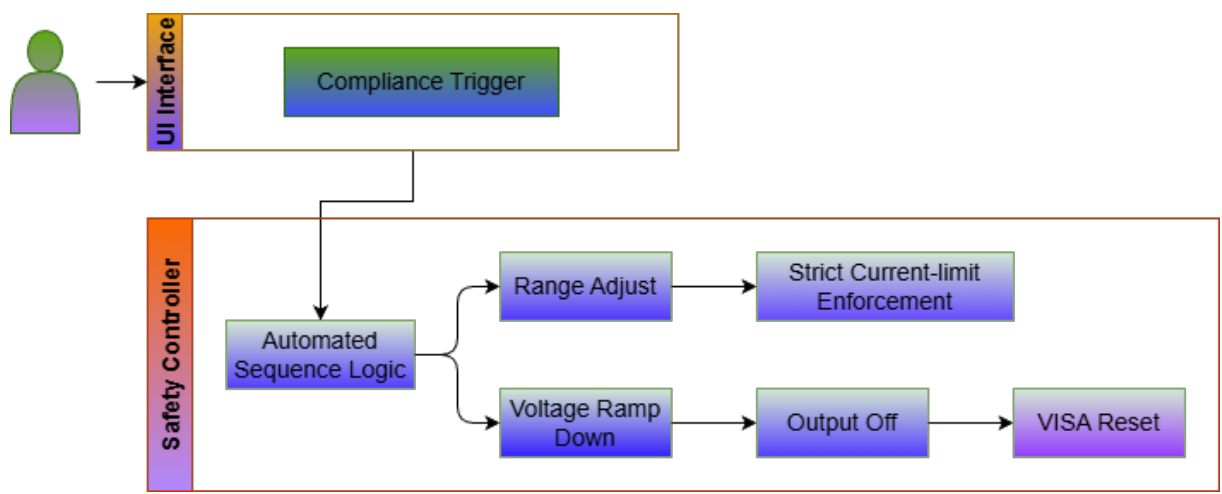

Figure 5. Safety and shutdown sequence illustrating compliance protection, automatic range adjustment, current-limit enforcement, voltage ramp-down, output shutdown, and VISA reset.

### 3.3. Electrochemical(EC) Sequence Control GUI

The EC-sequence control interface provides programmable waveform generation and synchronized streaming for memristive experiments. The GUI adopts a two-panel layout: the control frame manages user inputs---including sequence type, protocol parameters, and instrument settings---while the preview frame displays a pre-experiment waveform plot generated before any signal is applied to the device. This preview enables users to visually verify the shape and polarity of voltage or current sequences such as triangle, sine, square, to perform various memrisive experiments like PPF, SRDP, LTP/LTD or STDP/tSTDP before execution.

A parameterized sequence library exposes these protocols and also supports file-based sequence import for reproducible playback. Each generator is dynamically linked to a form composed of editable text fields and drop-down menus, allowing rapid switching between modes without code modification. Instrument control is fully integrated in the same window: VISA resources can be scanned on demand, and users specify compliance current and dwell-per-sample time before initiating output. The controller supports single-channel (A or B) and dual-channel operation with polarity mapping (A=+, B=− or A=−, B=+); when only one waveform is defined, the GUI can automatically synthesize an inverted counterpart to maintain symmetry tests.

Once verified, the selected sequence is streamed from a background worker thread to the Keithley 2600-series TSP driver, or via SCPI on non-TSP instruments, while a thread-safe acquisition callback logs per-sample metadata (index, t, $V_{set}$, $V_{meas}$, $I_{meas}$). Real-time acquisition and visualization are handled by a dedicated pop-up live-plot window, updated asynchronously through the same producer–consumer model as in Fig. 4. The acquisition thread continuously pushes measurement records into a bounded queue, and a lightweight consumer drains up to 20 frames $s^{-1}$ to update the plot without disturbing the output timing. This decoupling ensures that data streaming and GUI rendering remain fully non-blocking and timing-accurate even during long-duration runs.

After the sequence completes, an independent analysis window automatically opens, allowing the user to examine the recorded dataset in multiple coordinate modes, I-t, I-V, or I vs. sample index, and export all records to CSV for post-processing. Safety mechanisms reuse the compliance enforcement, range-transition handling, and multi-step shutdown path shown in Fig. 5, thereby safeguarding both the device under test and the instrument during unattended or high-frequency pulsing. Together, the EC-sequence GUI provides a unified environment for waveform design, pre-experiment preview, synchronized execution, and real-time logging, achieving reproducible neuromorphic pulse protocols with precise timing and hardware-protected operation.

### 3.4. Nanochannel Data Analysis GUI

The Analysis GUI provides a standalone environment for post-processing and visualising nanochannel datasets. It runs independently of the measurement interfaces while remaining fully compatible with their output formats, so that users can load I-V data either directly from the acquisition GUIs or from pre-existing files in .csv, .xlsx, or .txt format. Before any plots are drawn, the software applies an automated preprocessing stage that uses the physical parameters of the nanochannel (width, height, effective length) together with the bulk conductivity of the electrolyte. These inputs are used both to normalise the experimentally extracted conductance values and to compute the corresponding theoretical conductance or conductivity for quantitative comparison.

After preprocessing, the GUI performs data cleaning, applies linear regression to estimate conductance, and automatically locates the zero-current crossing point $E_{total}$. It also supports region-of-interest (ROI) based conductance extraction and context-aware unit conversion across I-V, G-V, and σ-C plots. Interactive plotting tools, including hover-based readouts, zooming, and annotations, are built in to facilitate detailed inspection of experimental trends.

A dedicated drift-diffusion and mobility module then uses these reduced quantities to compute the ion mobilities µ⁺ and µ⁻, as well as their ratio µ⁺/µ⁻, for both monovalent and more general salt models, providing deeper insight into charge-transport mechanisms within confined channels. All derived numerical results and figures can be exported for inclusion in laboratory records, publications, or for systematic benchmarking of experimental data against theoretical predictions. To make this computational step more broadly accessible, a lightweight web-based ion mobility calculator is also provided, allowing users to perform the same analysis without installing the full local software stack.

Beyond these analytical functions, the GUI includes a Reference Window that collates all equations, theoretical models, and their associated literature sources. Each entry lists the full citation, a DOI link, and a short summary of the underlying physical assumptions, so that users can trace every stage of the calculation back to the original publications. To add a small human touch to the otherwise technical workflow, a "Cheer-Up" module displays brief motivational messages when the program starts or closes. Taken together, these features allow the Analysis GUI to close the loop between acquisition, theory, and quantitative interpretation, transforming raw I-V datasets

### 3.5. Theoretical Framework

This section summarises the physical models and mathematical relations embedded within the software, which underpin the conductance, conductivity, and ion-mobility calculations. These formulations are not rederived here but are implemented in the Analysis GUI to enable direct comparison between expnd theoretical results.

#### 3.5.1 Theoretical Conductance Models:

The calculation of conductance depends on the geometry of the channel and the bulk ionic conductivity of the electrolyte. For a zero-dimensional nanopore, the total resistance combines channel and access resistances, given by[6]:

$$G = \sigma \left( \frac{4l}{\pi d^2} + \frac{1}{d} \right)^{-1} \quad (1)$$

For ultrathin 2D membranes where the internal resistance is negligible, this simplifies to

$$G = \sigma d \quad (2)$$

The one-dimensional and two-dimensional limiting cases are given respectively by[6]:

$$G_{1D} = \sigma \frac{\pi d^2}{4l} \quad (3)$$

$$G_{2D} = \sigma \frac{wbn}{l} \quad (4)$$

where σ is the bulk conductivity, d the pore diameter, and w, b, n, and l represent the width, height, number of channels, and channel length, respectively.

In the Nanochannel Data Analysis GUI, Eq. (4) and its corresponding geometric transformation are implemented to compute the theoretical conductance from user-specified device dimensions and electrolyte properties. Within the Enter Electrolyte Details and Analysis module, users can input or import electrolyte conductivities to generate theoretical conductance and conductivity values for comparison with experimental data. The same interface also enables the calculation of experimental conductivity from measured conductance and device geometry, allowing direct normalization and quantitative benchmarking against theoretical predictions.

### 3.5.2 Mobility Calculation

The ionic conductivity of an electrolyte is determined by the mobilities and concentrations of its charge carriers according to[6]

$$\sigma = F(|z^+|\mu^+ C^+ + |z^-|\mu^- C^-) \tag{5}$$

where F is Faraday's constant, $z^\pm$ are the ionic valences, $\mu^\pm$ the electrophoretic mobilities, and $C^\pm$ the respective ion concentrations. In the drift-diffusion module, the zero-current potential obtained from the I-V curve ($E_{total}$) is corrected for redox asymmetry arising from unequal ion activities and concentrations. The redox potential is given by[6]

$$E_{\text{redox}} = \frac{RT}{F} \ln\left(\frac{\gamma_H C_H}{\gamma_L C_L}\right) \tag{6}$$

where R is the universal gas constant, T is the absolute temperature, F is Faraday's constant, and $\gamma_{H,L}$ and $C_{H,L}$ are the activity coefficients and ion concentrations on the high and low concentration sides, respectively. The effective membrane (or osmotic) potential driving ion drift across the channel is then calculated as

$$E_m = E_{\text{osm}} = E_{\text{total}} - E_{\text{redox}} \tag{7}$$

representing the electrochemical potential difference after compensating for redox effects. For general electrolytes with ionic valences $z^+$ and $z^-$, the ratio of cation to anion mobility is expressed as[11]

$$\frac{\mu^+}{\mu^-} = -\frac{z^+}{z^-}, \frac{\ln(C_H/C_L) - z^- F E_m/RT}{\ln(C_H/C_L) - z^+ F E_m/RT} \tag{8}$$

In the case of monovalent salts, this simplifies to the Henderson equation[12],

$$\frac{\mu^+}{\mu^-} = \frac{\ln(C_H/C_L) + F E_m/RT}{\ln(C_H/C_L) - F E_m/RT} \tag{9}$$

Within the Analysis GUI, these relationships are applied sequentially to extract $E_m$ from experimental I-V curves and calculate the mobility ratio $\mu^+/\mu^-$. By combining the conductivity relation with measured concentration data, the program further derives the absolute mobilities $\mu^+$ and $\mu^-$ for both monovalent and general salts, enabling direct quantitative comparison between experimental transport data and theoretical predictions.

### 3.5.3 Power Density and Energy Conversion Analysis

The Analysis GUI also implements a module for estimating the theoretical maximum power density of nanochannel membranes based on measured conductance and zero-current potential. Following the standard osmotic energy conversion model[36], the maximum osmotic power ($P_{osm}$) and corresponding power density ($P_{density}$) are calculated from I-V data and channel geometry as

$$P_{\text{osm}} = \frac{1}{4} G E_m^2 \tag{10}$$

$$P_{\text{density}} = \frac{P_{\text{osm}}}{A_{\text{eff}}} = \frac{1}{4} \frac{G E_m^2}{nhw} \tag{11}$$

where G is the measured conductance (S), n is the number of nanochannels, and h and w are the height and width of each channel, respectively. The numerator $\frac{1}{4} G E_m^2$ corresponds to the maximum osmotic power under a matched resistive load, while the denominator n h w gives the effective cross-sectional area $A_{\text{eff}}$.

Within the GUI, users can compute $P_{osm}$ and $P_{density}$ directly after fitting $E_m$ from the I-V curve, facilitating rapid comparison of experimental performance across different devices and electrolytes.

### 3.5.4 Geometric and Normalisation Methods for Area Analysis

WeIn addition to physical transport models, the Analysis GUI integrates mathematical algorithms to quantify and normalize enclosed areas in cyclic I-V or G-V measurements. These routines enable the characterization of hysteresis magnitude, its evolution across voltage cycles, and fair comparison between datasets with different absolute ranges.

Polygon area calculation (Shoelace Formula): to compute the area enclosed by a closed I-V or G-V loop, the GUI employs the classical Shoelace formula[37] defined as

$$A = \frac{1}{2} \left| \sum_{i=1}^{n} (x_i y_{i+1} - x_{i+1} y_i) \right| \tag{12}$$

where ($x_i$, $y_i$) denote the coordinates of the $i^{th}$ point along the ordered loop and ($x_{n+1}$, $y_{n+1}$) = ($x_1$, $y_1$) closes the curve. The sign of A indicates the direction of traversal (positive for counter-clockwise and negative for clockwise loops), while the GUI takes its absolute value for plotting and quantitative comparison. This geometric computation provides a direct, model-independent measure of the hysteresis area enclosed between the forward and reverse voltage sweeps, implemented through vectorized summation for computational efficiency.

Area normalization (Min--Max Scaling): to compare loop areas across devices, frequencies, or voltage ranges, a min-max normalization is applied to rescale each area into a dimensionless quantity ranging from 0 to 1[38]:

$$A_{\text{norm}} = \frac{A - A_{\text{min}}}{A_{\text{max}} - A_{\text{min}}} \tag{13}$$

where $A_{\text{min}}$ and $A_{\text{max}}$ are the minimum and maximum areas within the dataset. This transformation preserves the relative magnitude and ordering of hysteresis sizes, ensuring consistency across measurements taken under different experimental conditions or scaling factors. Within the GUI, the normalized areas are plotted together with the corresponding I-V sweep frequencies, enabling visualization of how hysteresis magnitude evolves as a function of frequency and facilitating cross-device comparison under varied measurement rates.

Together, these mathematical modules support quantitative analysis of loop shape and stability, complementing the physical transport equations and mobility models described earlier.

## 4. Experiment and Results

The following subsections describe the experimental configuration and representative measurements obtained with the system. Fig. 6 summarises the I--V test workflow implemented in the Measurement GUI.

### 4.1. Experiment and Hardware Integration

This section describes the hardware and measurement configuration used to acquire the validation datasets summarised in Section B, and to test the GUIs under realistic operating conditions. The experimental platform is built around a Keithley 2600-series SourceMeter Unit (SMU), which provides programmable voltage sourcing and low-noise current measurement. The SMU is connected to a desktop PC via GPIB or USB-TMC, and the Python Tkinter GUIs communicate with the instrument through PyVISA using SCPI or TSP commands. During validation runs, the GUI configures sweep parameters, streams voltage and current data in real time, and renders live I-V plots using a non-blocking update loop to keep the interface responsive. All measurements, together with timestamps, channel identifiers, and configuration metadata, are logged to standard CSV or XLSX files for repeatable post-run analysis, reloading, and export in the Analysis GUI. Devices under test include nanochannel or nanopore chips and memristive nanochannel samples, mounted in a custom flow cell or probe station to ensure stable mechanical and electrical contact. For ionic measurements, Ag AgCl electrodes are used as reference and counter electrodes to minimise drift and provide reproducible potential referencing.

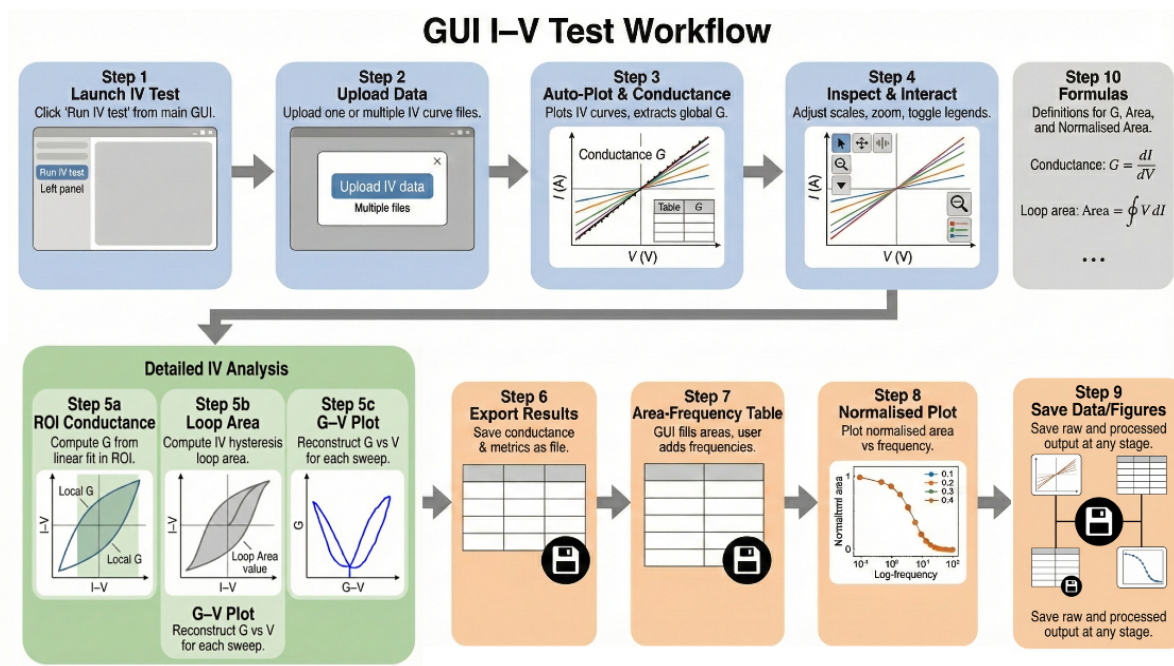

Figure 6. GUI I--V test workflow in the Measurement GUI. Starting from the main interface, the user launches an I-V test, uploads one or multiple I-V files, performs global and local conductance fitting, loop-area and G-V analysis, and exports summary tables and normalised area-frequency plots, all within a single interface.

## 4.2. Overview of Validation Datasets

The developed GUIs were validated using several complementary classes of datasets rather than a single experiment. First, a set of in-house nanochannel measurements provided multiple I-V curves, conductance-versus-concentration sweeps, and drift–diffusion measurements. These data were acquired on graphene and hBN nanochannel devices using aqueous electrolytes KCl, and form the basis for the multi-curve I-V analysis, conductance–concentration trends, and mobility examples presented in this section. The corresponding laboratory protocols and device configurations are summarized in Supplementary Section S1.

Second, the mobility calculator was benchmarked against an independent nanopore dataset digitized from the study on single-layer $MoS_2$ nanopores used to generate power[39]. This dataset provides well-characterized osmotic-power and ionic-transport behaviour, and allows us to verify that the GUI reproduces published mobility values when supplied with the same experimental parameters.

Third, power-versus-frequency and G-V validation were carried out using hysteretic I-V loops[13].

Finally, synthetic test signals and simulated voltage sweeps were used to probe timing performance, live-plot stability, and robustness of unit conversion and data parsing. These sequences exercise the acquisition loop and visualization pipeline under controlled conditions without relying on a specific physical device.

Due to space limitations, the main text focuses on four representative validation tasks: multi-curve I-V analysis, conductance–concentration analysis, mobility extraction, and area–frequency visualization, together with timing performance of the live plotting.

## 4.3. I–V processing and conductance–concentration(G–C)

To validate the integrated I-V and conductance-concentration workflow, we used a single, well–characterized graphene nanochannel and recorded KCl I-V curves over concentrations from $10^{-6}$ to 1 mol/L. All measurements were performed in a standard two–reservoir cell with Ag/AgCl electrodes, as sketched in Fig. 7(a). Once the raw data files are selected, the GUI handles the entire chain from file import, unit handling and fitting through to the final conductance-concentration plot, without requiring intermediate spreadsheet or plotting software.

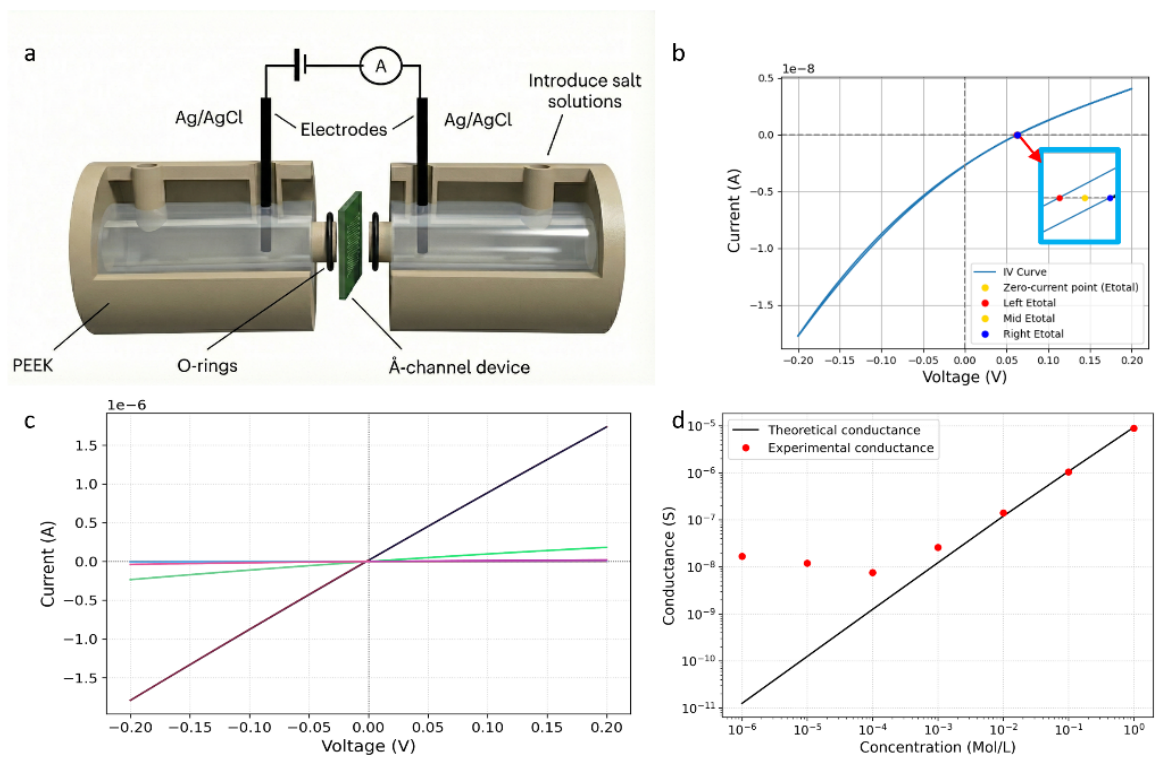

**Figure 7.** Experimental configuration and GUI-based analysis for an example nanochannel device. (a) Nanochannel cell with Ag/AgCl electrodes and Å-channel device. (b) I-V analysis window used to determine the total potential $E_{total}$. (c) Multi-curve I-V overlay and conductance extraction across KCl concentrations. (d) Conductance versus KCl concentration with GUI-derived experimental values and a theoretical line.

Fig. 7(b) illustrates the first stage of the workflow for a single I-V curve. The drift-diffusion analysis window displays the trace, performs a linear regression around zero bias, and automatically identifies the zero-current crossing (total potential $E_{total}$) together with left and right crossing points. The user can zoom, pan and read out exact voltage-current coordinates interactively, while the fitted slope and $E_{total}$ values are written directly into an editable results table that is reused by the drift-diffusion and mobility modules. This step replaces the previous practice of manually fitting each curve in a separate program and retyping the extracted parameters.

The second stage is the multi-curve analysis shown in Fig. 7(c). Here, a full concentration series of the example nanochannel device's I-V curves is loaded into a single window. The GUI overlays all traces with consistent voltage and current units, applies the same fitting procedure to each curve, and stores the resulting conductances in an internal table. Routine actions such as adding or removing curves, switching between multi-curve and single-curve views, toggling legends and saving both figures and tables are handled through dedicated buttons, so that exploring different subsets of the dataset does not require editing raw files.

In the final stage, the conductances are combined with the known geometric parameters of nanochannel device to compute ionic conductivity and to assemble the conductance-concentration trend. Figure 7(d) shows the KCl conductance as a function of concentration, with GUI-derived experimental points overlaid on a theoretical line based on bulk electrolyte properties. All red points and the black reference line in this panel are generated directly by the GUI from the same imported I-V files, demonstrating a smooth, human-centred workflow from raw current traces to publication-ready summary plots.

### 4.4. Drift--diffusion I--V and mobility analysis

For this experiment, a ten-fold concentration difference was applied across the nanochannel, producing a weakly nonlinear but monotonic I-V response around zero bias. A representative trace, together with the intermediate analysis steps, is shown in Supplementary Fig. S3.

For each drift-diffusion dataset, the I-V curve was first inspected in a narrow voltage range around the zero-current region to determine the total potential $E_{total}$. Three closely spaced estimates of the zero-current crossing were obtained from neighbouring data points, and the mid value was adopted as the effective membrane potential $E_m$. A low-bias interval was then selected on the same I-V curve, and a local linear regression was performed to extract the corresponding conductance G under the imposed concentration gradient.

The pair ($E_m$, G), together with the known concentration ratio across the nanochannel and the temperature, were inserted into the mobility expressions described in Section 3 to obtain the cation/anion mobility ratio $\mu^+/\mu^-$ and the individual mobilities.

To verify the Henderson-based mobility implementation against published data, we reanalysed the single-layer $MoS_2$ nanopore measurements reported in Ref.[39]. In these experiments, a freestanding $MoS_2$ membrane with a single nanopore separates two KCl reservoirs of concentrations $C_L$ and $C_H$, creating a chemical potential difference that drives an osmotic ion flux.

Fig. 8 shows how an I-V trace digitised from Fig. 3a of Ref.[39] is processed. The GUI simultaneously fits the measured short-circuit branch $I_{sc}$ (solid line) and the osmotic branch $I_{os}$ (dashed line). The zero-current intercept of $I_{sc}$ defines the total membrane potential $E_{total}$, while the intercept of $I_{os}$ gives the redox potential $E_{redox}$ of the Ag/AgCl electrodes at the two concentrations. Their difference is taken as the osmotic potential that enters the selectivity and mobility formulas. This workflow mirrors the physical picture in the original paper but packages the bookkeeping in a single, interactive step so that the user does not need to repeat the decomposition manually for each concentration pair.

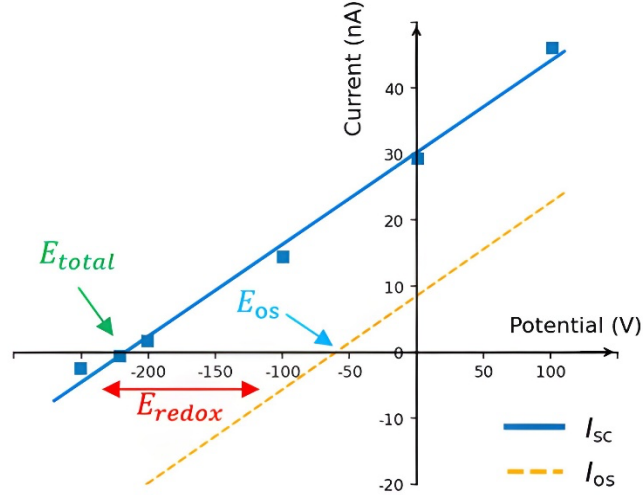

Figure 8. GUI-based analysis of an I-V trace digitised from Fig. 3a of Ref.[39]. The solid line corresponds to the measured short-circuit current $I_{sc}$, the dashed line to the osmotic branch $I_{os}$. Their linear fits define the total membrane potential $E_{total}$ (zero-current crossing of $I_{sc}$), the redox potential $E_{redox}$ (zero-current crossing of $I_{os}$), and the osmotic contribution $E_m$.

For a given asymmetric electrolyte configuration, the selectivity S associated with a concentration gradient and osmotic potential $E_m$ is[40]

$$S = \frac{E_m}{\left(\frac{RT}{F}\right)\ln\left(\frac{C_H}{C_L}\right)} \qquad (14)$$

or the same conditions, the cation/anion mobility ratio $\mu^+/\mu^-$ is obtained from the Henderson relation in (9). Conversely, the selectivity can be expressed directly in terms of the mobility ratio as[40]

$$S = \frac{\mu^+/\mu^- - 1}{\mu^+/\mu^- + 1} \qquad (15)$$

Mobility verification using single-layer MoS$_2$ nanopores: In our implementation, the literature values of $E_m$ and the digitized I-V curves from Ref. [39] are processed as in Fig. 8 to obtain $E_m$, which is then inserted into the above expressions together with the known concentration ratios to compute S and $\mu^+/\mu^-$. The resulting selectivities are summarized in Fig. 9(a). For four out of the five concentration ratios, the GUI values (blue squares) agree with the selectivities reported in Ref. [39] (orange open circles) to better than 1% relative error. The small discrepancies visible in the lower panel of Fig. 9(a) are consistent with uncertainties in the assumed room temperature and with the rounding of physical constants and tabulated potentials.

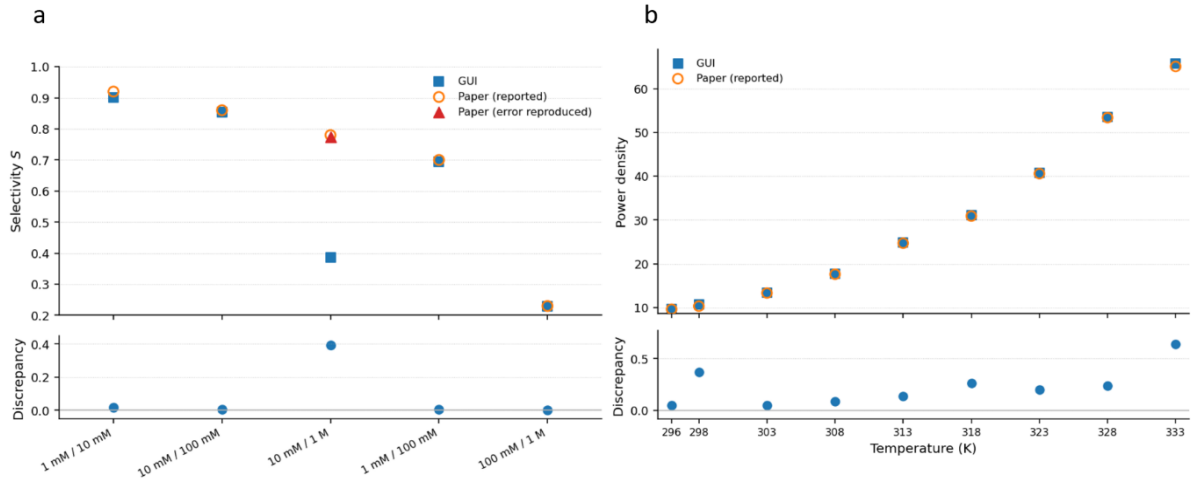

Figure 9 Combined verification of the mobility and power-density modules. (a) Selectivity S as a function of concentration ratio for single-layer MoS$_2$ nanopores, comparing GUI values (blue squares) with those reported in Ref.[39]; red triangles reproduce the inconsistent outlying value obtained with an incorrect concentration ratio. The lower panel shows the relative discrepancy between GUI and reported selectivities, which remains below 1% for all non-outlying points. (b) Maximum areal power density as a function of temperature for vermiculite Å-scale capillary arrays, comparing GUI calculations with the values of Rathi et al.[40]. The lower panel shows the small residual discrepancies, at the level expected from rounding of potential, R, and physical constants in the published data.

Power-density verification using vermiculite Å-scale capillaries: To validate the power-density calculator, we benchmarked against the vermiculite Å-scale capillary-array data of Rathi et al.[40]. For a like-for-like comparison, we matched their diffusion potential ($E_m$ = $V_{diff}$) and effective membrane area (nwh = 528 µm$^2$).

Using the diffusion potentials and resistances reported at different temperatures, we then input $E_m$ = $V_{diff}$ and G = 1/R into the "Power Density Calc" panel of the GUI (with the same geometric area) and compute P_max for each temperature. For the 296 K dataset ($V_{diff}$ ≈ 0.077 V and G ≈ 3.44×10$^{-6}$ S), the GUI returns $P_{max}$ = 9.65 W m$^{-2}$, in excellent agreement with the 9.6 W m$^{-2}$ value reported in Ref.[40]. Repeating this procedure across all temperatures reproduces both the exponential increase of power density with temperature and the reported ~578% increase from 296 K to 333 K. The comparison is shown in Fig. 9(b): the GUI results and literature values nearly overlap, and the residuals remain at the sub-percent level, consistent with rounding of the reported parameters.

### 4.5. Power-Frequency Analysis of Programmable Memristors

To verify that the loop-area based power-frequency analysis is applicable beyond our nanochannel devices. Figure 10(a) shows a schematic of the two-dimensional nanochannel memristor geometry used throughout this work, with Ag/AgCl electrodes driving ionic currents through an Å-scale slit in a SiN membrane. Panels b-d display families of hysteretic I-V curves digitized from three representative devices: the crossing 1 memristor from Supplementary Fig. 8B in Ismail et al.[13], and the two unipolar devices from Supplementary Fig. 12B and Fig. 12D in Robin et al.[41]. For each frequency and each device, the corresponding loop was imported into the GUI and the enclosed area was computed using the same polygon-based algorithm that is applied to nanochannel I-V traces. The loop area is proportional to the energy dissipated per cycle and, at fixed drive amplitude, provides a power-like figure of merit.

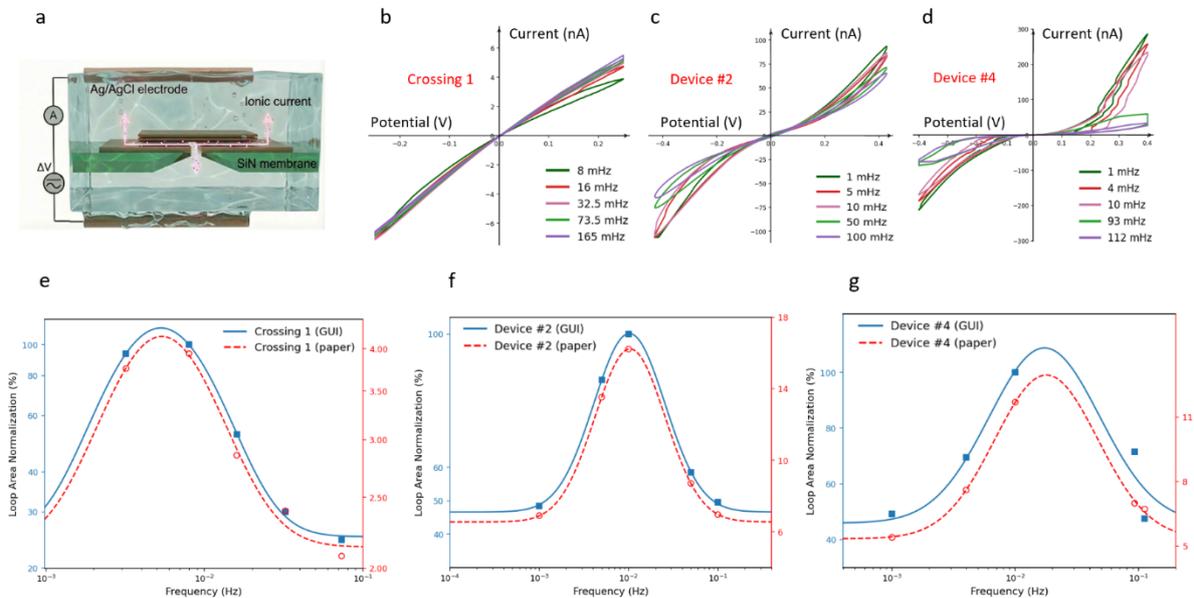

Figure 10. Verification of the loop-area based power-frequency analysis using published programmable nanofluidic memristors. (a) Schematic of the nanochannel memristor geometry and measurement configuration. (b-d) Families of I-V loops digitised from Refs.[13,41] for different devices and drive frequencies. (e-g) Comparison between normalised loop areas computed with our GUI (blue symbols and solid lines) and the memory-retention or normalised-area curves reported in the corresponding supplementary figures (red open symbols and dashed lines).

In the bipolar memristor of Ismail et al., the supplementary analysis (their Fig. 8A,B) identifies two conductance levels, $G_{on}$ and $G_{off}$, together with a symmetric voltage window from $-V_0$ to $+V_0$. In the idealized limit where the device switches abruptly between $G_{off}$ and $G_{on}$ at the turning points $\pm V_0$, the corresponding I-V hysteresis loop consists of two triangles whose total enclosed area is

$$A_{\text{ideal,bipolar}} = V_0^2 (G_{\text{on}} - G_{\text{off}}) \qquad (16)$$

The experimental loop area is obtained directly from the measured (or digitised) current-voltage characteristic as $A_{\text{loop,exp}} = \oint I(V)\, dV$, and the memory retention plotted in their Supplementary Fig. 8D corresponds to the normalised quantity:

$$A_{\text{norm}} = \frac{A_{\text{loop,exp}}}{A_{\text{ideal,bipolar}}} \qquad (17)$$

Panel e of Fig. 10 compares this bipolar normalization with the GUI output for the crossing 1 device. The blue symbols and solid curve show B(f), the percentage loop area obtained by feeding the digitized I-V loops of Supplementary Fig. 8B into the GUI, computing A_loop,exp(f) in the I-V plane, and normalizing it using the expression above. The red open symbols and dashed curve show the corresponding values C(f) extracted directly from Supplementary Fig. 8D. Both curves exhibit a single maximum at an intermediate frequency and reduced memory retention at lower and higher frequencies. The peak position and the characteristic width in frequency space are essentially identical, demonstrating that the GUI reproduces the frequency dependence of the bipolar normalization procedure of Ismail et al.

The devices in Robin et al.[41] can be treated in an analogous way, but with separate conductance levels on the positive and negative bias branches. In their Supplementary Fig. 12C,D the authors define four conductances, $G_{on,\pm}$ and $G_{off,\pm}$, evaluated at $\pm V_0$. The corresponding idealized I-V loop can be viewed as the sum of two triangular contributions, one from each bias polarity, giving the reference area

$$A_{\text{ideal,uni}} = \frac{1}{2} V_0^2 \left( (G_{on,-} - G_{off,-}) + (G_{on,+} - G_{off,+}) \right) \qquad (18)$$

and the normalised loop area is again calculated using equation (17). However, the absolute values of $G_{on,\pm}$, $G_{off,\pm}$ and $V_0$ are not reported numerically in Ref.[41]. As a result, $A_{\text{ideal,uni}}$ is known only up to an overall constant prefactor. In our validation (Fig. 10f,g), we therefore normalise the GUI-computed I-V loop areas are normalised by the maximum area across the measured frequencies as B(f) is defined as the experimental loop area $A_{\text{loop,exp(f)}}$ divided by the maximum value of $A_{\text{loop,exp(f)}}$ across all measured frequencies, and compare this to the published "normalised area" curves C(f) digitized from Supplementary Fig. 12B and Fig. 12D. The two definitions differ only by a constant scale factor set by the unknown value of $A_{\text{ideal,uni}}$, so any discrepancy in vertical scale reflects this missing prefactor, whereas the frequency dependence should coincide if the underlying normalisation scheme is consistent.

Panels f and g of Fig. 10 confirm that this is the case: for both unipolar devices the GUI-derived B(f) (blue) and the digitised C(f) (red) follow almost the same trend, with matching peak positions and similar asymmetry between the low- and high-frequency sides. Within the uncertainty associated with the missing numerical values of $G_{on,\pm}$, $G_{off,\pm}$ and $V_0$, the agreement in shape demonstrates that the loop-area computation and power-frequency workflow implemented in the GUI are fully compatible with the bipolar and unipolar normalisation procedures used in Refs.[13,41].

## 5. Conclusion

This work addresses the fragmentation between instrument control and data analysis in nanofluidic and memristive experiments by introducing a unified, Python-based GUI ecosystem. The framework connects three previously disjoint activities: (i) DC and pulsed I-V acquisition on nanochannels and nanopores, (ii) programmable waveform sequencing for neuromorphic and memristive tests, and (iii) physics-aware post-processing that converts raw traces into transport figures of merit such as conductance, mobility, osmotic power, and loop area. By bridging these domains, the system replaces ad-hoc, disjointed scripts with a cohesive workflow that enforces data integrity and experimental consistency from the moment of acquisition.

Architecturally, the system combines a Model-View-Controller software pattern with a producer-consumer threading model, allowing responsive Tkinter interfaces and live Matplotlib visualization without sacrificing timing regularity in the instrument loop. Safety routines, including automated compliance checks, range management, and soft ramp-down shutdown sequences, protect both sensitive nanofluidic devices and the SourceMeter hardware during unattended sweeps and high duty-cycle pulsing. This robust architectural design not only ensures operational safety but also establishes a reusable software template that can be extended to other instrument classes with minimal refactoring.

Experimental validation demonstrates that this integrated approach is not only convenient but also quantitatively reliable. In-house graphene nanochannel measurements on this device confirm that the Measurement and Analysis GUIs can take a full KCl concentration series from raw I-V curves to conductance-concentration and conductivity plots within a single workflow, reproducing expected bulk trends. Crucially, the automated analysis pipeline eliminates the potential for human error associated with manual data handling. Benchmarking against digitized literature datasets shows that the drift-diffusion module recovers ion selectivity and mobility ratios for single-layer $MoS_2$ nanopores to within reported uncertainties, while the power-density calculator reproduces the temperature-dependent osmotic power of vermiculite capillary arrays with near-perfect overlap. The EC-Sequence and hysteresis modules also capture the frequency-dependent loop areas of programmable nanofluidic memristors, matching both bipolar and unipolar normalization schemes used in recent studies. These results confirm that the embedded physical models possess the accuracy required for high-impact academic research.

Overall, the proposed GUI ecosystem provides a scalable foundation for next-generation nanofluidic and memristive experiments. By unifying acquisition, control, and physics-based analysis - accessible through both the comprehensive desktop environment and a lightweight web-based calculator - it lowers the barrier to quantitative transport studies, facilitates direct benchmarking against theoretical models and published data, and supports the development of more standardized, sharable workflows across the nanofluidic research community. Ultimately, this open-source framework serves as a step towards digitizing the complete lifecycle of nanofluidic experimentation, fostering greater transparency and reproducibility in the field.

**Code and Data Availability**

All Python source code for the graphical user interfaces and analysis workflows developed in this work is openly available in the Nanofluidic GUI Toolkit GitHub repository: https://github.com/YICHAOWANG131419/yichaowang-Nanofluidic-GUI-Toolkit. The repository includes the Nanochannel Analysis GUI, measurement and control routines for nanochannel, nanopore, and memristor experiments, user documentation, and both an archived HTML version and a live browser-based deployment of the ion mobility calculator: https://YICHAOWANG131419.github.io/yichaowang-Nanofluidic-GUI-Toolkit. All tools are compatible with Python 3.9 or later and rely only on standard scientific libraries (NumPy, SciPy, Matplotlib, Pandas, and Tkinter). Installation and usage instructions are provided in the repository documentation.

The source data required to reproduce the main figures of this paper are available in the Source data directory of the same GitHub repository. Additional raw measurement files not included in the public archive are available from the corresponding author upon reasonable request.

## Competing interest

The authors declare no competing interests.

## Acknowledgement

The author gratefully acknowledges Dr. R.K. Gogoi, Dr. M.S. Martins and Ms. Sangeeta from the Angstrofluidics group of University of Manchester for their valuable suggestions in the development of this software.

# Supplementary Information

# An Open-Source Framework for Measurement and Analysis of Nanoscale Ionic Transport

Yichao Wang[1], Munan Fang[1,2], Aziz Roshanbhai Lokhandwala[1,2], Siddhi Vinayak Pandey[1,2], Boya Radha [1,2]*

[1]Department of Physics and Astronomy, School of Natural Sciences, The University of Manchester, Manchester M13 9PL, United Kingdom

[2]National Graphene Institute & Photon Science Institute, The University of Manchester, Manchester M13 9PL, United Kingdom

*Correspondence to be addressed to: Boya Radha (email: radha.boya@manchester.ac.uk)

# Contents

# 1. Software Environment and Installation:

This section lists the software requirements and explains how to start the graphical user interfaces (GUIs).

**Python Version:**

All programs run with Python version 3.9 or later. The same Python version is used for all three GUIs.

**Analysis GUI:**

The analysis GUI is a Python program for reading and processing nanochannel and nanopore data on a desktop or laptop computer. The main Python packages are numpy, pandas, matplotlib, mplcursors and tkinter. PyVISA is not needed here, because this GUI only loads data files that were recorded earlier. To start the analysis GUI, the user opens a terminal in the analysis folder and runs the main analysis script (for example, analysis_gui.py). The window has two parts. The left panel contains file controls, unit conversion options, analysis buttons and export options. The right panel shows either the current plot or a results table, depending on the selected action.

**Measurement GUI:**

The measurement GUI controls a Keithley SourceMeter and records current-voltage data and memristor pulse sequences. The main Python packages are numpy, pandas, matplotlib, mplcursors and tkinter. PyVISA is required to communicate with a real instrument. Openpyxl or xlsxwriter are optional packages for saving results to Excel files. he measurement GUI also uses two small local modules. One module provides a background service for live plotting. The other module defines the window that shows the live trace.

If PyVISA is not installed, the program prints a warning when it starts and runs in an offline simulation mode. In this mode the GUI generates test current-voltage curves so that the interface can be used without a real device.

To launch the measurement GUI, the user runs the main control script in the measurement folder. The main window then allows the user to choose the SourceMeter channel, set the sweep range and current limit, start and stop measurements, and save the recorded data and figures.

# 2. GUI Layout：

Figures in this section show the main layouts of the three GUIs and the mobility analysis window. They are numbered with an "S" prefix (for example, Fig. S1) as recommended for supplementary documents.

**Measurement GUI Layout:**

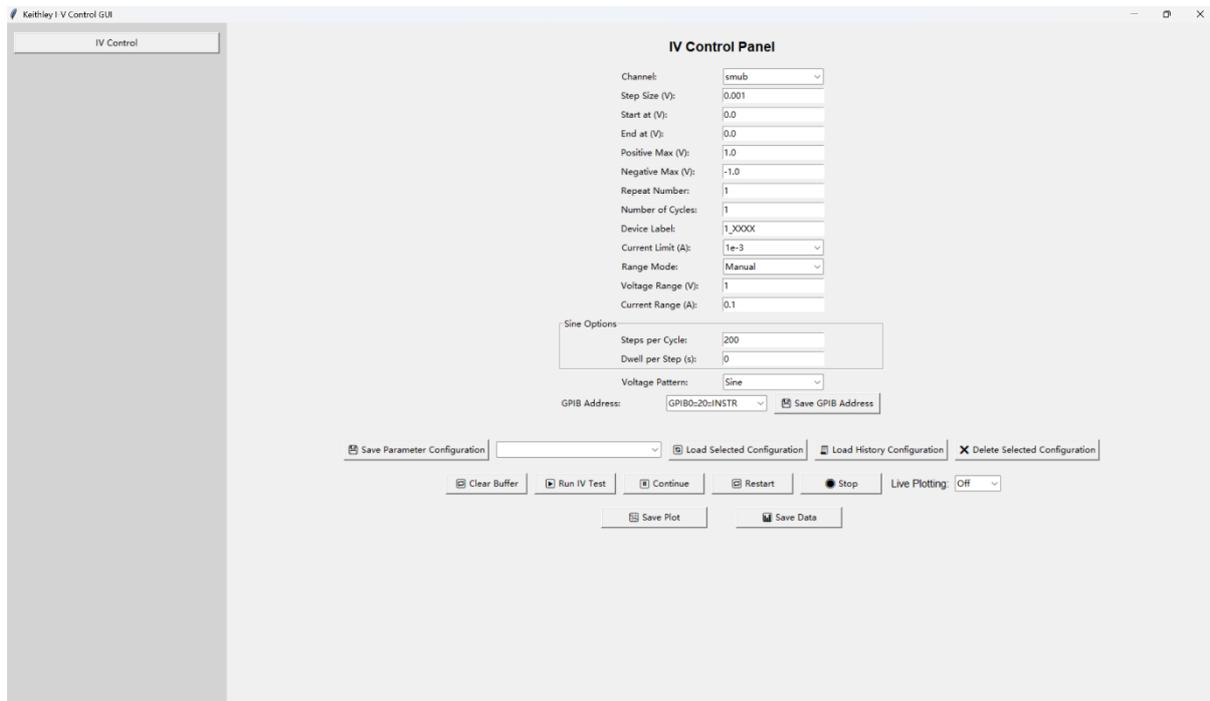

Fig. S1. Main control window of the Keithley measurement GUI used for I--V sweeps and memristor tests.

Figure S1 shows the main window of the Keithley measurement GUI. The window has three main parts. A narrow panel on the left contains the navigation button for the I-V control panel. The central area shows the I-V control panel itself. The bottom row contains buttons for configuration and data handling.

In the I-V control panel the user sets all parameters needed for a sweep. These fields include the channel, step size, start and end voltages, number of steps, positive and negative limits, repeat cycles, current limit, range mode and voltage range. There is also a box for the GPIB address and a button to store the address.

The bottom row of buttons is used to save and load parameter configurations, clear the buffer, start and stop an I-V test, enable or disable live plotting, and save the final plot and data files.

**EC Measurement GUI Layout:**

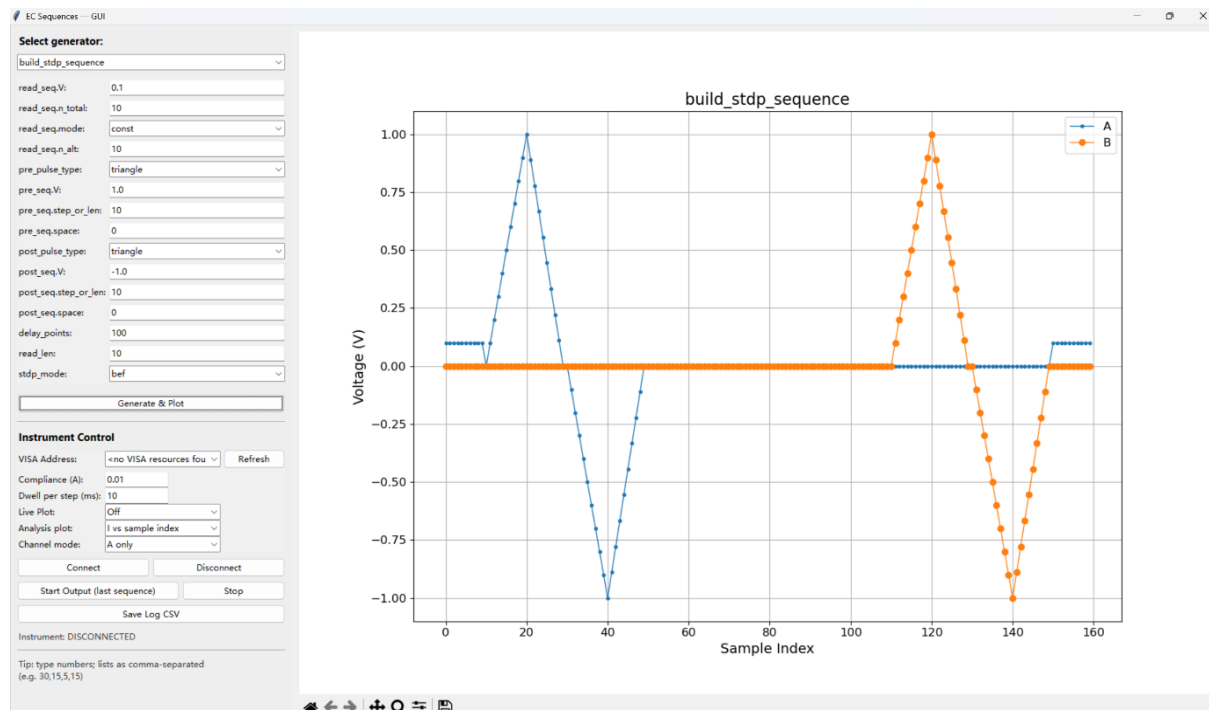

Fig. S2. Main window of the EC measurement GUI used to build and send STDP pulse sequences to the instrument.

Figure S2 shows the main window of the EC measurement GUI. The window is split into a left control panel and a right plot. On the left side, the upper block contains the pulse generator settings. Here the user selects the generator type and sets the parameters for the spike timing dependent plasticity (STDP) sequence, such as amplitudes, widths and delays for the A and B pulses. Below the generator settings there is an instrument control block. This block holds the VISA address, connection buttons, output mode and the controls for sending the sequence to the device and starting or stopping the test. The right side shows the generated voltage sequence as a function of sample index for the two channels. The plot updates when the user changes the parameters. A small toolbar at the bottom of the plot provides basic zoom, pan and save functions.

**Analysis GUI Layout:**

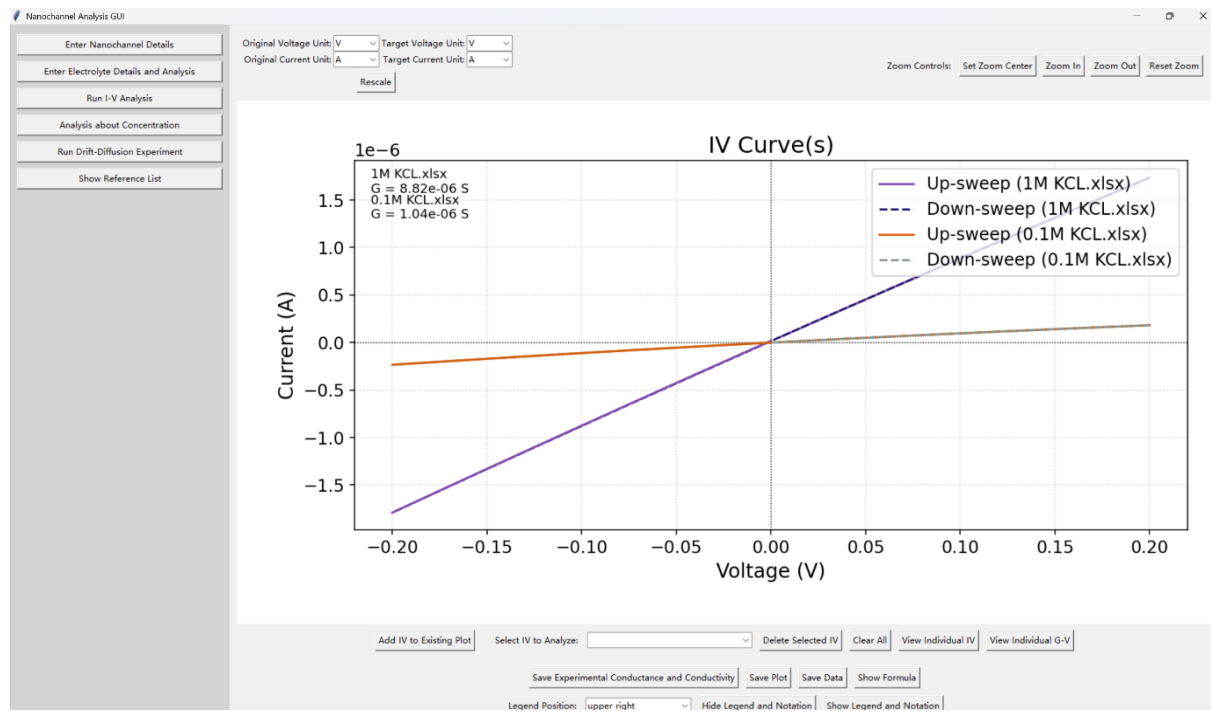

**Fig. S3.** Main window layout of the nanochannel analysis GUI used for multiple analysis.

Figure S3 shows the main window of the nanochannel analysis GUI. The left panel contains the main control buttons. From top to bottom, the user can enter nanochannel geometry, enter electrolyte details, run I-V analysis, perform concentration-based analysis, start the drift diffusion experiment module and open the reference list.

Across the top of the window there are drop down menus for the original and target voltage and current units, together with a Rescale button. A small group of zoom buttons on the right controls the view of the active plot.

The central area shows the current I-V plot. Multiple up sweep and down sweep curves can be displayed together, with a legend in the upper right corner. Fitted conductance values are printed directly on the plot for each loaded data set.

Along the bottom of the window there is a row of buttons for adding new I-V curves to the existing plot, selecting a curve for further analysis, deleting or clearing traces, viewing individual curves, saving plots and data, and showing the formula used for conductance and conductivity. A small drop-down menu sets the legend position and can hide or show the legend and text annotations.

## 3. Example Device Geometry:

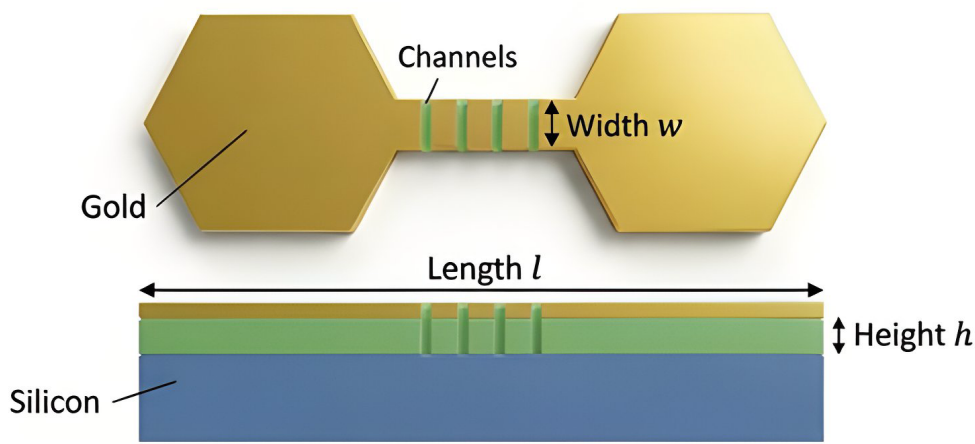

**Fig. S4.** Top and side view of the example nanochannel device. The narrow-slit region contains n parallel channels of height h, width w and effective length l between the gold pads.

Figure S4 shows a simple sketch of the example nanochannel device. The top view shows two gold pads connected by a narrow nanochannel region, and the side view shows the channel height on the silicon substrate. For this device used in this work the geometric values are l = 5.8 µm, w = 26 µm, h = 7.9 Å and n = 235, and these values are used in the power density and mobility calculations in the analysis GUI.

## 4. Published Data Used in Figures:

The following panels use previously published data digitised from the cited references. Fig. 8(b) uses an I-V trace digitised from Fig. 3(a) of Feng et al.[1]. Fig. 9(a) uses conductance and selectivity data digitised from Extended Data Fig. 3 ("Ideal cation selectivity of the pore") of Feng et al.[1]. Fig. 9(b) uses mobility and conductivity values taken from Table S2 of Rathi et al.[2]. Fig. 10(b)-(d) use hysteretic I-V loops digitised from Supplementary Fig. 8B of Ismail et al.[3] and Supplementary Figs. 12B and 12D of Robin et al.[4].

## 5. Code and Data Availability:

The Python source code for all GUIs and analysis scripts used in this work is available in the Nanofluidic GUI Toolkit repository on GitHub https://github.com/YICHAOWANG131419/yichaowang-Nanofluidic-GUI-Toolkit.

The repository includes the Nanochannel Analysis GUI, the measurement and control GUIs, the ion mobility calculator and example analysis scripts.